\documentclass[12pt,useAMS,usenatbib]{article} 
\usepackage{graphics}
\usepackage{graphicx} 
\usepackage{amsmath} 
\usepackage{jcappub}
\usepackage{natbib} 
\usepackage{hyperref} 
\usepackage{color}
\usepackage{psfrag} 
\usepackage{epsfig}

\title{\boldmath Constraining warm dark matter power spectrum using the
  cross-correlation of  HI 21 cm signal and the
  Lyman-$\alpha$ forest}

\author[a]{Anjan Kumar Sarkar, }
\author[b]{Ashis Kumar Pal, }
\author[b]{Tapomoy Guha Sarkar}
\affiliation[a]{Raman Research Institute, Bangalore, India.}
\affiliation[b]{Birla Institure of Technology and Science Pilani, Pilani-Campus, India}

\emailAdd{anjans@rri.res.in}
\emailAdd{p2017106@pilani.bits-pilani.ac.in}
\emailAdd{tapomoy1@gmail.com}

\abstract{ We have considered the prospects for measuring the cross Warm Dark Matter (WDM) power spectrum of the redshifted HI 21-cm signal and the Lyman-$\alpha$ forest and thereby constraining WDM mass using observations with upcoming radio-interferometers - the Ooty Wide Field Array (OWFA) and SKA1-mid, and a spectroscopic survey of the quasars. We have considered a quasar survey with a mean observed quasar number density of $\bar{{\rm n}}_{\rm Q} = 48$ deg$^{-2}$ over a collecting area of 14455 deg$^{2}$, and a mean spectroscopic SNR = 5. Our analysis with OWFA shows that it is possible to measure the WDM power spectrum in several $k$-bins at $k < 0.4$ Mpc$^{-1}$ with ${\rm SNR} > 5$ using an observation of 200 hours each in 100 different fields-of-view for $m_{\rm WDM} = 0.25$ keV. Considering the possibility of the joint measurement of the parameters, the warm dark matter density parameter $\Omega_{{\rm WDM}}$, and the dark energy density parameter $\Omega_{\Lambda0}$, we find that the relative error on the $1-\sigma$ measurement of the parameter $\Omega_{{\rm WDM}}$ is $\sim 0.8$ for a fiducial $m_{\rm WDM} = 0.25$ keV. We further find that it is possible to have a measurement of the suppression of power from the Cold Dark Matter (CDM) power spectrum at a confidence level of $\sim 7.2-\sigma$ and $\sim 2.7-\sigma$ in two different $k$-bins over the $k$-range $0.1 \leq k \leq 3.13$ Mpc$^{-1}$ for $m_{\rm WDM} = 0.15$ keV. Considering the analysis with SKA1-mid, we find that for a fiducial $m_{\rm WDM}= 0.25$ keV, the  suppression in the cross  power spectrum can be measured at $\sim 10 - \sigma$ around $k \sim 0.2 \rm{Mpc}^{-1}$ for a total observing time of $20000$ hrs distributed uniformly over $50$ independent pointings where the available $k$-range is binned as $\Delta k = k/5$. }

\begin{document}
\maketitle
\flushbottom
\section{Introduction}
\label{sec:intro}
In recent times the $\Lambda$CDM model has been widely accepted as the
standard model of cosmology with strong support from CMBR observations \cite{aghanim2018planck},
Galaxy surveys \cite{SDSS14} and other cosmological and astrophysical
probes. However, several fundamental questions remain unanswered.  While, one
is still unsure about the actual physical nature of cold dark matter, there is
a major discrepancy between the observed abundance of dwarf galaxies in the
local group in comparison to a far greater number predicted by CDM
simulations.
The number of dwarf galaxies observed in voids are also seen to be much lesser
than predictions from CDM models \cite{moore1999dark,klypin1999missing,peebles2010nearby,diemand2007formation}.
There has also been some difference between the observed shallow rotation
curves  and the ones obtained from CDM simulations which typically produces  a
cuspy inner density profile \cite{navarro1997universal,stadel2009quantifying}.
The proposal of a Warm dark matter (WDM) attempts  to explain some of these
discrepancies of the cold dark matter models. Dark matter particles with
velocities in the transition zone between relativistic and non relativistic
region are characterized as 'warm'. Mass of the WDM particle falls in region
between eV to GeV and are fiducially assumed to be in the keV range \cite{boyarsky2009lyman,boyarsky2008constraints,boyarsky2009realistic,seljak2006can}.
Gravitinos and sterile neutrinos are some of the proposed candidates for warm
dark matter \cite{ellis1984supersymmetric,dodelson1994sterile,seljak2006can}.  Unlike CDM, WDM can only
cluster on a scale greater then its Jeans scale.

Warm dark matter particles are known to remain relativistic at early
times. However their density fluctuations are suppressed owing to free
streaming on the scales which are comparable to the horizon size at those
epochs.  This leads to a consequent suppression of the power spectrum on small
scales. At later times the WDM particles undergo cooling due to cosmic
expansion and their late time behavior mimics the CDM with some residual
velocity dispersion \cite{dodelson1994sterile,bode2001halo,boyanovsky2011warm}. It is evident that lighter WDM particles shall
remain relativistic for longer time and thereby have a larger free-streaming
scale. Consequently the formation of halos of mass $M_{fs}$ with $M_{fs} \propto
({m_{_{\rm WDM}}})^{-\chi}$ shall be suppressed, where $\chi$ is some positive
parameter.  The free streaming of warm dark matter particles manifests through
the modification of the matter transfer function.

Intensity mapping of the collective HI 21-cm radiation emission from the post-reionization era 
is believed to provide
invaluable information regarding the large scale matter distribution, and
expansion history of the Universe \cite{BNS2001,bharadwaj-sethi2001,wyithe-loeb2008,bharadwaj2009,visbal2009}.
Several radio telescopes like the Giant Metrewave Radio Telescope\footnote{http://www.gmrt.ncra.tifr.res.in/}
(GMRT)\cite{ananthakrishnan1995GMRT}, the Ooty Wide Field Array (OWFA)\cite{prasad2011,subrahmanya2017OWFA}, the Canadian
Hydrogen Intensity Mapping Experiment\footnote{https://chime-experiment.ca/}(CHIME) \cite{bandura2014CHIME}, the Meer-Karoo Array
Telescope\footnote{https://www.ska.ac.za/gallery/meerkat/} (MeerKAT), and the Square Kilometer Array
(SKA)\footnote{https://www.skatelescope.org/} have dedicated goals towards detecting the cosmological HI
signal. The major challenge towards detecting the signal is however posed by large
galactic and extra-galactic foregrounds \cite{ghosh2011}. Several
other observational errors like calibration errors and man made radio
frequency interferences make it further difficult for the signal to be
detected. A statistical detection of the signal with high SNR involves very
careful noise analysis and subtraction of foregrounds \cite{ghosh2011,ghosh2012foregrounds,prasad2013foregrounds,ali-bharadwaj2014}.

The diffuse HI from the post reionization epoch also may be mapped out using
the distinct absorption features in Lyman-$\alpha$ forest, which traces out the
HI density fluctuations along one dimensional sight lines of background QSOs.
The Lyman-$\alpha$ forest observations are known to have numerous applications
in cosmological investigations like the  measurement of matter power spectrum
\cite{croft-weinberg1998lyman,croft-weinberg1999lyman,croft-weinberg2002lyman} and the bispectrum \cite{mandelbaum2003lyman,vielbispectrum2004lyman}, estimation of cosmological parameters \cite{mcdonald1999lyman,lesgourgues2007lyman}, constraining
reionization history \cite{gallerani2006reionization} and modelling of dark energy \cite{mcdonald2007lyman} etc.
Different sources of observational error pertaining to the Lyman-$\alpha$
observations arise mainly from improper modeling and subtraction of the
continuum, improper modeling and inclusion of the fluctuations of the
ionizing source, uncertainties in the 
temperature-density relation in the IGM  \cite{hui-gnedin1997,weinberg1997lymana,weinberg1997lymanb} and  metal line contaminations
\cite{kim2007metal}.
The Baryon Oscillation Spectroscopic Survey (BOSS) aims to use  the imprint of
BAO in the Lyman-$\alpha$ forest as a probe of dark energy.
The present catalog of SDSS \cite{SDSS14} indicates the availability of a large
number of QSO spectra with high signal to
noise ratio (SNR). This allows us to do a 3-dimensional analysis of the
Lyman-$\alpha$  forest and thereby improve the constraints on cosmological parameters.

Numerical simulations have revealed that on large cosmological scales both the
post reionization redshifted 21-cm signal and the Lyman-$\alpha$ forest are 
biased tracers of the underlying  matter distribution
\cite{bagla2010,guha-sarkar2012,sarkar2016,carucci2017}. The cross-correlation
of the Lyman-$\alpha$ forest and the redshifted 21cm signal from the
post-reionization epoch has been established as a potentially useful probe of
the cosmological power spectrum and several works have explored the
possibility of using this as a probe of the post-reionization Universe \cite{Guha-sarkar2011cross,guha-sarkar2015,pal2016,guha-sarkaretal2016,guha-sarkarsen2016,sarkar2018cross}. The
cross-correlation signal has been ascertained by both linear analysis \cite{guha-sarkar2015} and robust numerical
simulations \cite{carucci2017}.  The cross-correlation technique has been proposed to be
a way to bypass some of the major observational issues \cite{guha-sarkar2015}. There also has
been the proposal of cross-correlating the 21-cm signal with the Lyman break
galaxies \cite{villaescusa2015lymanbreak}. A successful detection of the HI 21-cm emission at redshift
$z \sim 0.8$ using cross correlations of HI 21-cm maps and galaxies has been
reported \cite{chang2010HIintensity}. The foregrounds which plagues the 21cm observations
are expected to pose less severe challenges in detecting the cross-correlation
signal as the the foregrounds in HI 21-cm observations appear only as a noise
in the cross correlation and can therefore be tackled for a statistically
significant detection.

In this paper we investigate the possibility of measuring warm dark matter mass through
the way it affects the cosmological power spectrum. We  consider the 3D cross
power spectrum of the post reionization HI 21-cm signal  and
the large scale Lyman-$\alpha$  forest. We discuss the possibility of
detecting the cross-correlation signal in a WDM cosmology  using future Lyman-$\alpha$ forest
surveys with very high QSO number densities and  two radio telescopes - the OWFA and the  upcoming SKA-mid phase1
(SKA1-mid). These two radio interferometers are chosen for our analysis since they have distinctly
different array layouts and observational parameters.
We make predictions for warm dark matter masses and the possibility of
statistical detection of the suppression effect of WDM  on the binned cosmological power spectrum.

\section{The redshifted HI 21-cm and  the Lyman-$\alpha$ forest cross-correlation signal in a WDM cosmology}

Warm dark matter suppresses the growth of perturbations on scales that are smaller than the free streaming scale $\lambda_{WDM}$. The free streaming scale is found to be inversely related to the WDM mass $m_{WDM}$ as $\lambda_{WDM} \propto m_{WDM}^{-4/3} \Omega_{WDM}^{1/3}$ (which corresponds to a mode $k_{WDM} = 2 \pi / \lambda_{WDM}$) \cite{smith2011WDM}. This would lead to an erasure of structures of masses smaller than
 
\begin{equation}
\frac{4}{3} \pi \lambda_{WDM}^3 \bar{\rho} 
\end{equation}

where $\bar{\rho}$ is the mean background density. The free streaming scale introduces a modification to the CDM matter power spectrum through a suppression in the matter transfer function. The transfer function in the WDM model is related to the CDM transfer function as 

\begin{equation}
T_{WDM}(k) = \left  [ 1 + (\alpha k)^{2\mu} \right ]^{-10/\mu} T_{CDM}(k)
\end{equation}

where the parameters are obtained from numerical simulations \cite{viel2005constraining} as $\mu = 1.12$ and $\alpha$ is given by 

\begin{equation}
\alpha = 0.049 \left [\frac{m_{WDM}}{keV} \right]^{-1.11}\left [ \frac{\Omega_{WDM}}{0.25} \right]^{0.11}\left [\frac{h}{0.7} \right]^{1.22}h^{-1} Mpc
\end{equation}

We use the linear transfer function from \cite{eisenstein-hu1998} to compute the WDM power spectrum using the above fit function. The suppression of scales smaller than the free streaming scale, if detected, shall allow us to measure WDM mass. The halo model based non-linear WDM transfer function is discussed in the Appendix. 

Following the complex phase transition during the epoch of reionization \cite{loeb2001reionization,barkana2001reionization,mesinger2015reionization}, most of HI in the post-reionization era ($z \leq 6$) is believed to be clumped in the highly dense regions that are identified as the Damped Lyman Alpha (DLA) systems in quasar observations. The redshifted 21 cm radiation from individual HI clouds is very weak. However, radio  observations in the  frequency range $210 {~\rm MHz} \leq \nu_{\rm HI} \leq {1420}{~\rm MHz}$ holds the potential to tomographically map out the collective diffuse emission from these systems in the post-reionization era. 

The CMBR brightness temperature changes from $T_{\gamma}$
to $T(\tau_{_{21}})$ under radiative transfer through a HI cloud at redshift $z$ along the line of sight ${\bf \hat{n}}$. This is due to the emission or the absorption associated with the 
the spin flip Hyperfine transition of HI in its rest frame at frequency $\nu_c = 1420 \, {\rm MHz}$.
The primary quantity of interest in a radio-interferometric observations is the excess brightness temperature $T_b ({\bf\hat{n}}, z)$ that is written as,
  
\begin{equation}
 T_b( {\bf\hat{n}}, z)
= \frac{T(\tau_{_{21}}) - T_{\gamma}}{1 + z} \approx \frac{(T_s -
  T_{\gamma})\tau_{_{21}}}{1 + z}.
\end{equation}

at a redshift $z$, $\tau_{_{21}}$ gives the HI 21-cm optical depth.
 
The fluctuations in $T_b( {\bf\hat{n}}, z)$ is given by 
   $\delta_T(r{\bf\hat{n}}, z) = \bar T(z) \times \eta_{\rm
  HI}(r{\bf\hat{n}}, z)$, where $r$ is the comoving distance corresponding to
$z$, 
\begin{equation}
\bar T(z)=4.0 {\rm mK} (1+z)^{2}\left(\frac{\Omega_{b0}h^{2}}{0.02}\right)\left(\frac{0.7}{h}\right) \left( \frac{H_{0}}{H(z)}\right)
\end{equation}
and
\begin{equation}
\begin{split}
\eta_{HI}( r {\bf\hat{n}}, z)=\bar{x}_{HI}(z) \left\{ \left ( 1 - \frac{T_\gamma}{T_s}\right)\left[\delta_{H}(z,{\bf\hat{n}})-\frac{1+z}{H(z)}\frac{\partial v}{\partial r}\right] \right. \\ 
\left.+
 \frac{T_\gamma}{T_s} s \delta_{H}({\bf\hat{n}}r,z) \right\}
\end{split}
\end{equation}
Here $ \bar{x}_{HI}(z) $ is the mean neutral fraction, $
\delta_{H}(z,{\bf\hat{n}})$ is the density fluctuations in the HI and the function $s$ relates the fluctuations of the spin temperature with that of the HI density \cite{bharadwaj-ali2004CMB}. The peculiar velocity of the gas, $v$ leads to the  anisotropic term $(1+z)/ H(z)\frac{\partial v}{\partial r}$.

The post reionization epoch is characterized by  $T_{\gamma}/T_s << 1 $ owing to rapid rise of $T_s$ at low redshifts, and the 21 cm signal is seen in emission. We then have,  
\begin{equation}
\eta_{_{HI}}( r {\bf\hat{n}}, z)=\bar{x}_{_{HI}}(z)\left[\delta_{H}(z,{\bf\hat{n}})-\frac{1+z}{H(z)}\frac{\partial v}{\partial r} \right].
\end{equation} 
The fluctuation  $\delta_{ T} (\textbf{r})$ in Fourier space is denoted by 
$\Delta_T (\textbf{k})$ and is given by \cite{guha-sarkar2015}
\begin{equation}
\Delta_{T}(\textbf{k})=C_{T}[1+\beta_{T}\mu^{2}]\Delta(\textbf{k})
\end{equation}
where $\Delta(\textbf{k})$ is the Fourier transform of the underlying dark matter over density $\delta$.  The peculiar velocity of the gas is assumed to sourced solely by dark matter overdensity leading to redshift space distortion which is  quantified through the parameter  $\beta_T$  and $ \mu = {\bf\hat{n}} \cdot  {\bf\hat{k}}$. 
The quantity $C_T(k, z) = \bar T(z) \bar{x}_{HI}(z) b_T(k, z)$ gives the amplitude of the fluctuation, where the bias $b_T(k, z)$ relates the  HI
fluctuations $\Delta_H(\textbf{k})$  to dark matter
fluctuations $\Delta(\textbf{k})$ through $\Delta_{{\rm HI}} (\textbf{k})  =  b_T(k,z) \Delta (\textbf{k})$. Apart from the cosmological parameters, the post-reionization HI is essentially modeled using two functions
$\bar{x}_{HI}(z)$ and  $b_T(k, z)$. 

The post-reionization HI bias has been extensively studied using numerical simulations \cite{bagla2010,guha-sarkar2012,sarkar2016} . Most of these simulations rely on some canonical way to populate the haloes with neutral hydrogen and consequently identify them as DLAs. The HI in halos should have some minimum threshold circular velocity so that it may shield itself from ionizing radiation. This threshold sets a lower bound for the halo mass $M_{\rm min}$. Further, very massive halos also do not contain any HI \cite{pontzen2008lyman}. The total neutral gas is  distributed
to halos within a chosen mass range such that the mass of the gas assigned to a halo is proportional to the mass of the halo.

Using this simple scheme it has been found that the HI bias grows monotonically with $k$ on small scales.
Some additional scale dependence of the bias is also owes its origin to the fluctuations in the ionizing background.
On large scales, the  bias is  however found to be a constant increasing only with redshift.
Noting that our model assumes that bulk of the neutral gas is contained in halos, cosmologies with massive neutrinos and warm dark matter show a greater HI clustering than the model with only cold dark matter. This is because matter fluctuations at smaller scales are wiped out due to the free streaming effect of the warm dark matter and the neutrinos, whereby smaller mass halos are rarer in these models.

The Lyman-$\alpha$ forest traces out the small fluctuations in the HI density in the largely ionized IGM
along the line of sight to distant quasars where they manifests as a distinct absorption features in the observed quasar spectra. Whereas the 21-cm signal in the post reionization era is sourced by the dense DLA clouds, the Lyman-$\alpha$ forest is sourced by the tiny HI fluctuations in the predominantly ionized IGM. The transmitted QSO flux through the Lyman-$\alpha$ forest is given by the fluctuating Gunn-Peterson effect as 

\begin{equation}
\mathcal{F} = \bar{\mathcal{F}} e^{- A ( 1 + \delta)^\Gamma }
\end{equation}

where $\bar{\mathcal{F}}$ denotes the mean transmitted flux, $\Gamma$ is a parameter dependent on the slope of the temperature-density power law relation, and the parameter $A \sim 1$ has implicit dependence on the astrophysical properties of the IGM and other cosmological parameters.
However, on a reasonably smoothed scale, the fluctuation in the transmitted flux $\delta_{\mathcal{F}} = (\bar{\mathcal{F}} - \mathcal{F})/\bar{\mathcal{F}} \propto \delta$.
This linear dependency on large scales has been studied and validated by numerical simulations of the Lyman-$\alpha$ forest \cite{carucci2017}. 

In a manner similar to the HI 21-cm signal, the Fourier space fluctuations in the transmitted flux of the Lyman-$\alpha$ forest can be written as,

\begin{equation}
\Delta_{\mathcal{F}} ({\bf k})= C_{\mathcal{F}} ( 1 + \beta_{\mathcal{F}}  \mu^2 ) \Delta ({\bf k}).
\end{equation}

The parameter $\beta_{\mathcal{F}}$ quantifies the anisotropy in $\Delta_{\mathcal{F}} ({\bf k})$ in a manner similar to the redshift space distortion parameter of the HI 21-cm signal $\beta_T$. However, the non-linear relation between the Lyman-$\alpha$ transmitted flux and the underlying dark matter density field makes the interpretation of $\beta_{\mathcal{F}}$ different from that of $\beta_{\rm T}$ in that $\beta_F$ is not to be treated as the bias parameter for the Lyman-$\alpha$ forest. Further, $C_T$ and $\beta_T$ are independent parameters, but are both dependent on the HI bias $b_T$, whereas $C_{\mathcal{F}}$ and $\beta_{\mathcal{F}}$ has no such common factor. It has been seen in numerical simulations that fluctuations in the Lyman-$\alpha$ flux can be well described by a linear theory with a scale independent bias on large scales.

We express the  three dimensional power spectrum of 21-cm signal, the Lyman-$\alpha$ forest and the cross correlation
generally as

\begin{equation}
\langle \Delta_a({\bf k}) \Delta_b^{*}({\bf k'}) \rangle = (2 \pi)^3 \delta^3({\bf k}- {\bf k'}) P_{ab} ({\bf{k}})
\end{equation}

where $a$, $b$ can generally be $\mathcal{F}$ and $T$. In redshift space, the expression for $P_{ab} ({\bf{k}})$ is is given by

\begin{equation}
P_{ab}({\bf k}) = C_a C_b(1 + \beta_a \mu^2)(1 + \beta_b \mu^2) P(k)
\label{eq:pftk}
\end{equation}

where $\mu = k_{\parallel}/k$, the direction cosine of the wave-vector to the line-of-sight, and $P(k)$ gives the matter power spectrum (this is $P^{\rm WDM}$ in our analysis). The auto-correlation power spectrum corresponds to $a = b$ and the cross-correlation power spectrum corresponds to $a = T$ and $b = \mathcal{F}$.

We have used $C_{\mathcal{F}} = -0.13$ and $\beta_{\mathcal{F}} = 1.58$ for the Lyman-$\alpha$ forest at redshift $z_c = 2.55$ in our analysis. This is obtained from the fit to the 1-D Lyman forest power spectrum \cite{palanque2013}. For the HI 21-cm signal, we have used $C_T = (x_{\rm HI} b_{\rm HI} \bar{T}$ and $\beta_T =  f(\Omega)/b_{\rm HI}$, where $x_{\rm HI}$, $b_{\rm HI}$, $f(\Omega)$, and $\bar{T}$ respectively are the mean neutral hydrogen fraction, HI bias, linear growth rate of the matter density perturbations, and the characteristic brightness temperature for the HI 21-cm signal \cite{ali-bharadwaj2014,bharadwaj-ali2005}. We have used $A_c = C_{\mathcal{F}} C_T$ as the amplitude of the cross correlation power spectrum, and $A_T = C_T^2$ and $A_F = C_{\mathcal{F}}^2$ give the amplitude of the power spectrum of the HI 21-cm signal and the Lyman-$\alpha$ forest respectively.

We have calculated the mean neutral hydrogen fraction $x_{\rm HI}$ using the relation $x_{\rm HI} = \Omega_g/\Omega_b$ where $\Omega_b$ and $\Omega_g$ refer to the baryon density parameter and the neutral gas density parameter in the universe respectively. DLA observations \cite{zafar2013,prochaska2009,noterdaeme2012} have measured $\Omega_g \sim 10^{-3}$ over a redshift range $1 < z < 5$. This corresponds to $x_{\rm HI} = 0.02$, which we have used in our analysis. Semi-numerical simulations of the post-reionization HI 21-cm signal \cite{bagla2010,villaescusa2014modelling} are found to be consistent with a scale-independent, linear HI bias at large scales ($k < 1 \, {\rm Mpc}^{-1}$). However, HI bias becomes highly non-linear as we go down to smaller scales ($ k > 1 \, {\rm Mpc}^{-1}$). We have accounted for this behavior by using a scale and redshift dependent HI bias \cite{sarkar2016}. The value of $f(\Omega)$ has been calculated using the $\Lambda$WDM cosmological parameters given in \cite{aghanim2018planck}. 

\section{Some observational aspects}

We have used the quasar number distribution from the DR14 of SDSS \cite{SDSS14}.
The quasar distribution is seen to peak at $z = 2.25$, and falls off as we
move away from the peak. It is worthwhile to consider the cross-correlation at redshifts near to the peak. This shall enable us to accommodate a large number QSO sightlines thereby reducing the noise contribution arising from the discrete sampling of the QSOs. For the purpose of the present analysis, we have chosen a fiducial redshift of $z_c = 2.55$. For a quasar at $z_Q$, we note that we eliminate the part of the spectra $10, 000$ km s$^{-1}$  blue-ward of the Lyman-$\alpha$ emission peak to avoid the the quasar proximity effect  and
also consider the part of the spectra that is  beyond $1,000$ km s$^{-1}$
red-ward of the Lyman-$\beta$  line or the O-VI lines to avoid the confusion of  the Lyman-$\alpha$ forest with other absorption lines. 

For the given quasar, there is a restrictive redshift range
for which the quasar spectrum maybe used for cross-correlation. The
cross-correlation is also only possible in the region of overlap between this
redshift range and the band width of the  21-cm observation. We have considered both complete and partial overlap to estimate the mean quasar number density $n_Q(z_c)$. For the Lyman-$\alpha$ forest, the actual signal to noise ratio (SNR) can be as large as 10. We have adopted an uniform value of $SNR = 5$ for our analysis.
 
The discussion till now has been restricted to 21-cm observations in a single pointing
direction. Typically, the field of view of the radio interferometer  is much smaller than the area covered
by spectroscopic surveys like BOSS, and it is worthwhile to also consider the possibility
of extending the analysis to a situation where 21-cm observations are carried out in
multiple  pointing directions. In the present work, we assume the cross-correlation signal from each pointing direction to be statistically independent of each other whereby the Fisher matrix for the combined observation is the sum of individual fisher matrices for each pointing directions. It is important to note that for carrying out the cross-correlation, both the Lyman-$\alpha$ and the HI 21-cm signal are to be smoothed at the same resolution. Given that both the observations shall have different frequency resolutions, we have smoothed both the signals at the coarser resolution amongst the two. 

\section{Results}

\begin{figure}
\begin{center} 
\psfrag{T1}[c][c][0.75][0]{$200 \times 25$ hours \quad \quad \quad \quad \quad \quad}
\psfrag{T2}[c][c][0.75][0]{$200 \times 50$ hours \quad \quad \quad \quad \quad \quad}
\psfrag{T3}[c][c][0.75][0]{$200 \times 100$ hours \quad \quad \quad \quad \quad \quad}
\psfrag{x1}[c][c][0.8][0]{$k \, {\rm Mpc}^{-1}$}
\psfrag{y1}[c][c][0.8][0]{SNR}
\vskip.2cm
\centerline{{\includegraphics[scale=0.75]{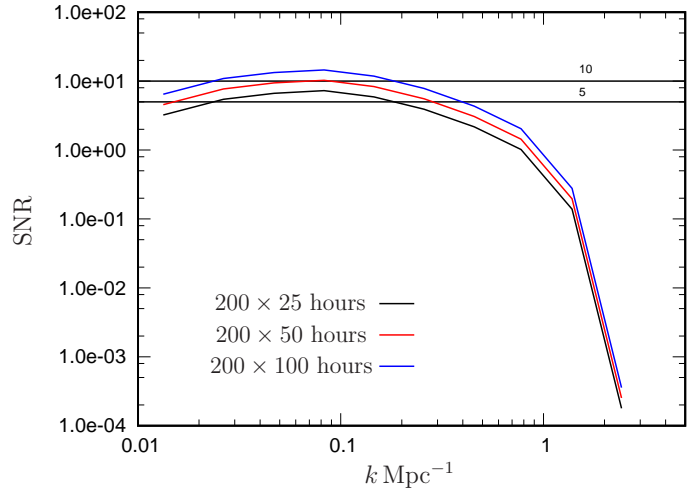}}}
\caption{Shows the SNR for measuring the WDM cross power spectrum $P_{\mathcal{F}T}^{\rm WDM}(k)$ in different $k$-bins with observations of 200 hours each in $N_{\rm p}$ = 25, 50 and 100 different fields-of-view. We have used a value of $m_{\rm WDM}$ = 0.25 keV for this analysis. The lower and upper horizontal lines in the figure correspond to SNR = 5 and 10 respectively.}
\label{fig:snrbin}
\end{center}
\end{figure}

\subsection{ Predictions for the Ooty Wide Field Array (OWFA)}

The OWFA is a linear radio-interferometric array that is
expected to operate a central frequency of $\nu_c = 326.5 \, {\rm MHz}$ (or, an
wavelength of $\lambda_c = 0.9135$ m). This corresponds to observing the
HI 21-cm radiation from a redshift $z_c = 3.35$
\citep{prasad2011,marthi2013}. The OWFA is a $530$m long and $30$m wide
parabolic cylindrical reflector that is placed along the north south
direction on a hill at a slope of 11$^{\circ}$, which is equal to the
latitude of the place \citep{swarup1971,sarma1975}. This makes it possible to
track a given part of the sky using a single rotation of the telescope about
the telescope's long axis. The OWFA feed system consists of 1056
half-wavelength ($\sim 0.5\lambda_c$) dipoles, spaced $0.48$ m equally apart,
placed almost end-to-end along the long axis of the cyllinder. OWFA can
operate in two independent simultaneous radio-interferometric modes - PI and
PII \citep{ali-bharadwaj2014}. The PI and PII respectively have $40$ and $264$
antennas in total, corresponding to the situations where signals from $24$
dipoles and $4$ dipoles have been added to make an single antenna element
respectively. For the purpose of our analysis, we have only considered OWFA
PII. The PII has the smallest and the largest baselines of $1.92$ m and
$505.0$ m respectively. Both PI and PII have an operating bandiwdth of 39 MHz. 

The possibility of detecting the HI 21-cm signal using OWFA has been studied
extensively \citep{ali-bharadwaj2014,gehlot-bagla2017OWFA,marthi2017OWFA,chatterjee2017OWFA,chatterjee2018OWFA}. Detailed foreground predictions \citep{ali-bharadwaj2014,marthi-chatterjee2017OWFA} and  calibration issues \cite{marthi-chengalur2013OWFA} for OWFA have also been addressed.

\subsubsection{The WDM power spectrum estimation using a visibility based approach to the cross-correlation}

We begin our analysis by considering the possibility of constraining the shape
of the cross power spectrum directly from future observations. To this end, we have
assumed that the values of $\beta_T$ and $\beta_F$ are known a priori, and
have considered constraining the shape of the cross power spectrum
$P_{\mathcal{F}T}(k)$ (eq.~\ref{eq:pftk}) using observations of the cross-correlation
signal with OWFA and an spectroscopic survey like SDSS-IV. 

The prospects of measuring the binned cross power spectrum for the redshifts, $z_c = 3.35, \, 3.05 \, {\rm
  and} \, 2.55$, and for the observing bandwidths, $B = 30 \, {\rm and} \, 60
\, {\rm MHz}$ has been studied in an earlier work \cite{sarkar2018cross}. The study shows that we have the best possible measurement prospects of the binned cross power spectrum for the redshift and
the bandwidth of 2.55 and 60 MHz respectively. In this work, we have considered observing
the cross correlation signal at a redshift of 2.55 that corresponds to HI
observation at a frequency of 400 MHz, and with a observing bandwidth of 60
MHz. Given this frequency and bandwidth, OWFA PII covers the $k$-range $0.010
\leq k \leq 3.13 \, {\rm Mpc}^{-1}$. We have used a system temperature of $T_{{\rm sys}} = 100$ K to calculate the  noise variance (eq. 3.4 in \cite{sarkar2018cross}) in our analysis. 

As studied earlier in \cite{sarkar2018cross}, the SNR for detecting the cross power
spectrum grows rather slowly for observing time beyond 200 hours in a single
field-of-view. This indicates that the SNR for observing time beyond 200 hrs in a 
single field-of-view is dominated by the cosmic variance. It is therefore reasonable to consider carrying out observation of 200 hours each in ${\rm N}_p$ different independent pointing directions
whereby the total observation time is, T = $200 {\rm N}_p$. We have carried
out our analysis with three different observing times, T = 5000, 10000 and
20000 hours that respectively correspond to observing in ${\rm N}_p = 25, \,
50 \, {\rm and} \, 100$ independent fields-of-view. For the purpose of the present analysis, we have binned the OWFA PII $k$-range into 10 equally spaced logarithmic $k$-bins. We have adopted the visibility based approach developed in \citep{} to study the prospects of detecting the WDM power spectrum using the cross correlation of Lyman-$\alpha$ forest and HI 21-cm signal with an upcoming radio-interferometric array OWFA and an spectroscopic survey like BOSS. We made the noise estimates using eqs. (3.10 - 3.14) in \cite{sarkar2018cross}.  

Figure \ref{fig:snrbin} shows the predicted SNRs for measuring the cross power
spectrum (eq.~\ref{eq:pftk}) in different $k$-bins for $m_{\rm WDM} = 0.25$ keV. 
We expect the measurement errors to be
dominated by the cosmic variance at small $k$ whereas at large
$k$, the errors are predominantly due to the system noise.We
find that it is possible to have a measurement of the cross power spectrum
with ${\rm SNR} \geq 5$ in a number of bins within the range $0.02 \leq k \leq
0.2 \, {\rm Mpc}^{-1}$ with an observation of 200 hours each in ${\rm N_p} =
25$ different fields-of-view. For observations with ${\rm N_p} = 50$
fields-of-view, we find that it is possible to have a measurement with ${\rm
  SNR} \geq 5$ for a number of bins within the range $0.015 \leq k \leq 0.25 \,
{\rm Mpc}^{-1}$. Measurement with an SNR in excess of 10 is possible in
a single bin centred at $k = 0.08 {\rm
  Mpc}^{-1}$. Prospects improve further if we consider observation with even
more fields-of-view, ${\rm N_p} = 100$, where it is possible to have a
measurement with ${\rm SNR} \geq 5$ for a number of $k$-bins at $k < 0.4 \,
{\rm Mpc}^{-1}$. Measurement with ${\rm SNR} \geq 10$ is
even possible in three $k$-bins within the range $0.02 \leq k \leq
0.15 \, {\rm MPc}^{-1}$. The results do not vary significantly if we consider
carrying out our analysis with smaller $m_{\rm WDM}$ values, $m_{\rm WDM}$ =
0.20, 0.15 and 0.10 keV. We here emphasize that in the limit where the SNR is 
dominated by the cosmic variance, the noise is roughly proportional to the 
signal itself, whereby the SNR remains insensitive to the signal. 

\begin{figure}
\begin{center} 
\psfrag{c1}[c][c][0.8][0]{$1-\sigma$ \quad }
\psfrag{c2}[c][c][0.8][0]{$2-\sigma$ \quad }
\psfrag{c3}[c][c][0.8][0]{$3-\sigma$ \quad }
\psfrag{c4}[c][c][1][0]{$m_{\rm WDM}$ = 0.25 keV \quad \quad \quad \quad \quad \quad}
\psfrag{y1}[c][c][1][0]{$\Delta \Omega_{\rm WDM} / \Omega_{\rm WDM}$}
\psfrag{x1}[c][c][1][0]{$\Delta \Omega_{\Lambda 0} / \Omega_{\Lambda 0}$}
\vskip.2cm
\centerline{{\includegraphics[scale= 0.80]{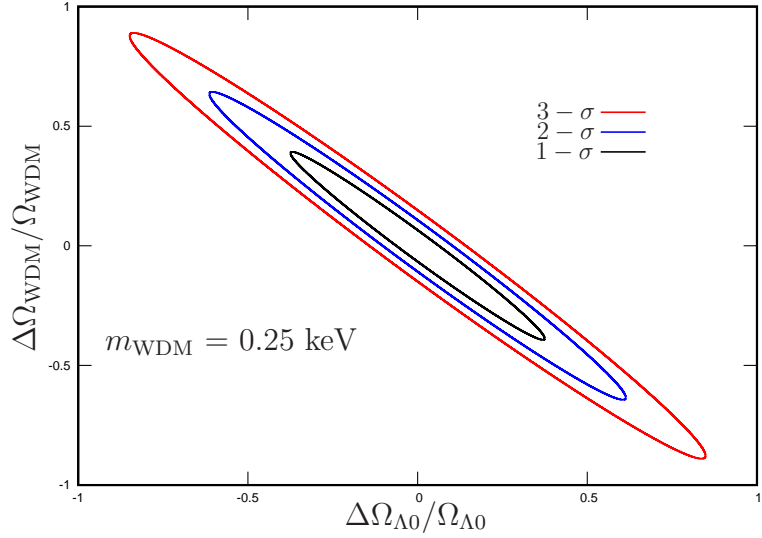}}}
\caption{Shows the relative $1-\sigma$, $2-\sigma$ and $3-\sigma$ errors on the joint measurement of $\Omega_{\rm WDM}$ and $\Omega_{\Lambda 0}$ with an observing time of 200 hours each in $N_{\rm p}$ = 100 different fields-of-view for $m_{\rm WDM}$ = 0.25 keV.}
\label{fig:error}
\end{center}
\end{figure}

We now consider the prospects of the joint measurement of the two parameters,
$\Omega_{\rm WDM}$, the warm dark matter density parameter, and
$\Omega_{\Lambda 0}$, the dark energy density parameter marginalizing over the
amplitude of the cross power spectrum. We have considered observations
of the cross correlation signal for 200 hours each in 100 different
fields-of-view. Figure \ref{fig:error} shows the relative errors in the joint
measurement of $\Omega_{\rm WDM}$ and $\Omega_{\Lambda 0}$ for $m_{\rm WDM}$ =
0.25 keV. We see that the errors in the measurements of the parameters are
anti-correlated. The relative errors in the measurement of both the parameters
are roughly the same, the relative $1-\sigma$, $2-\sigma$ and $3-\sigma$
measurement errors are respectively $\sim$ 0.4, $\sim$ 0.6 and $\sim$ 0.8. We
have also considered carrying out the analysis with $m_{\rm WDM}$ = 0.20 and 0.15 keV where we find that the errors increase slightly as the value of
$m_{\rm WDM}$ is decreased.

\begin{figure}
\begin{center}
\vskip.2cm 
\psfrag{T1}[c][c][0.6][0]{\quad \quad \quad \quad \quad \quad $m_{\rm WDM}$ = 0.10 keV} 
\psfrag{T2}[c][c][0.6][0]{\quad \quad \quad \quad \quad \quad $m_{\rm WDM}$ = 0.15 keV}
\psfrag{T3}[c][c][0.6][0]{\quad \quad \quad \quad \quad \quad $m_{\rm WDM}$ = 0.20 keV} 
\psfrag{binned wdmpowerspectrum}[c][c][0.4][0]{binned WDM cross power spectrum \quad \quad \quad \quad \quad \quad \quad \quad}
\psfrag{binned neutrinopowerspectrum}[c][c][0.4][0]{binned matter cross power spectrum \quad \quad \quad \quad \quad \quad \quad \quad \quad } 
\psfrag{wdmpowerspectrum}[c][c][0.4][0]{WDM cross power spectrum \quad \quad \quad \quad \quad \quad }
\psfrag{neutrinopowerspectrum}[c][c][0.4][0]{Cross power spectrum \quad \quad \quad \quad \quad  }
\psfrag{PFT}[c][c][0.65][0]{$P_{\mathcal{F}T}(k)$ mK Mpc$^3$}
\psfrag{k}[c][c][0.65][0]{$k$ Mpc$^{-1}$}
\centerline{{\includegraphics[scale =.45]{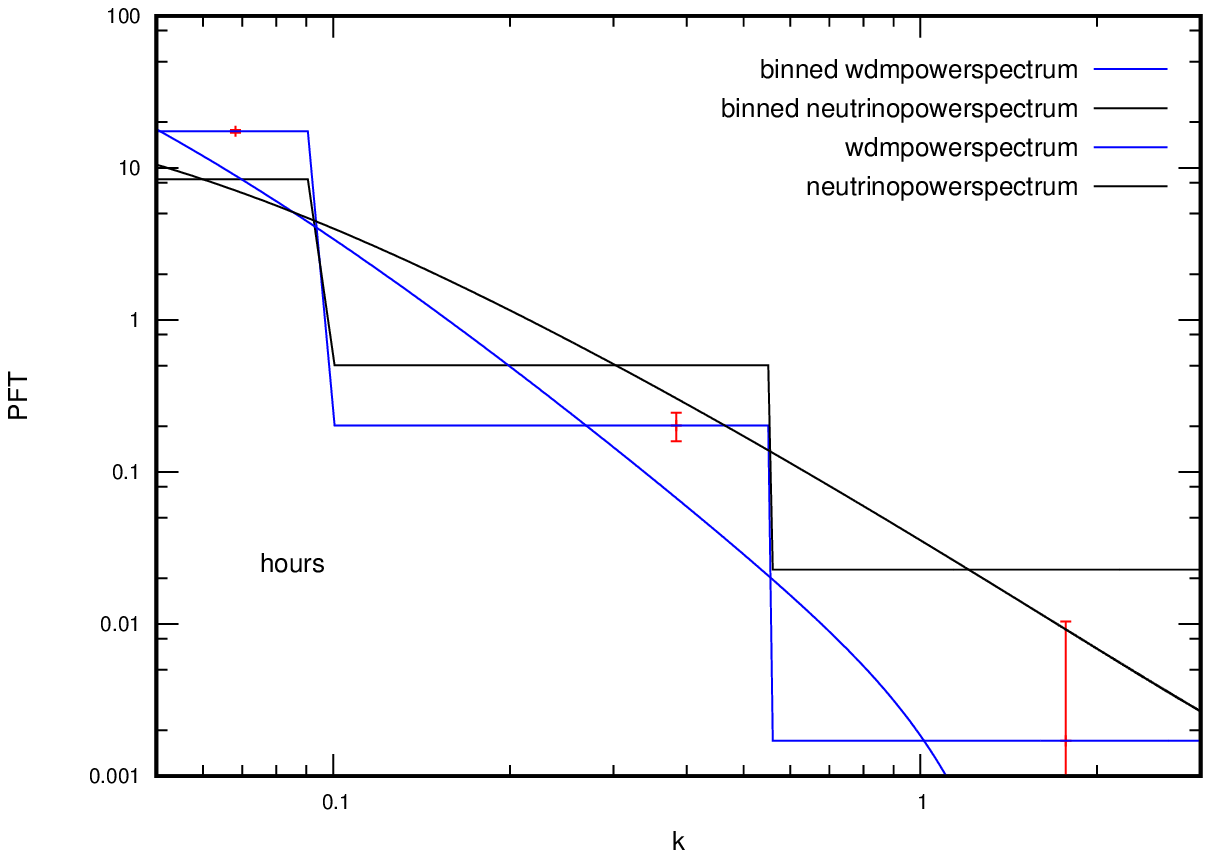}}
  {\includegraphics[scale =.45]{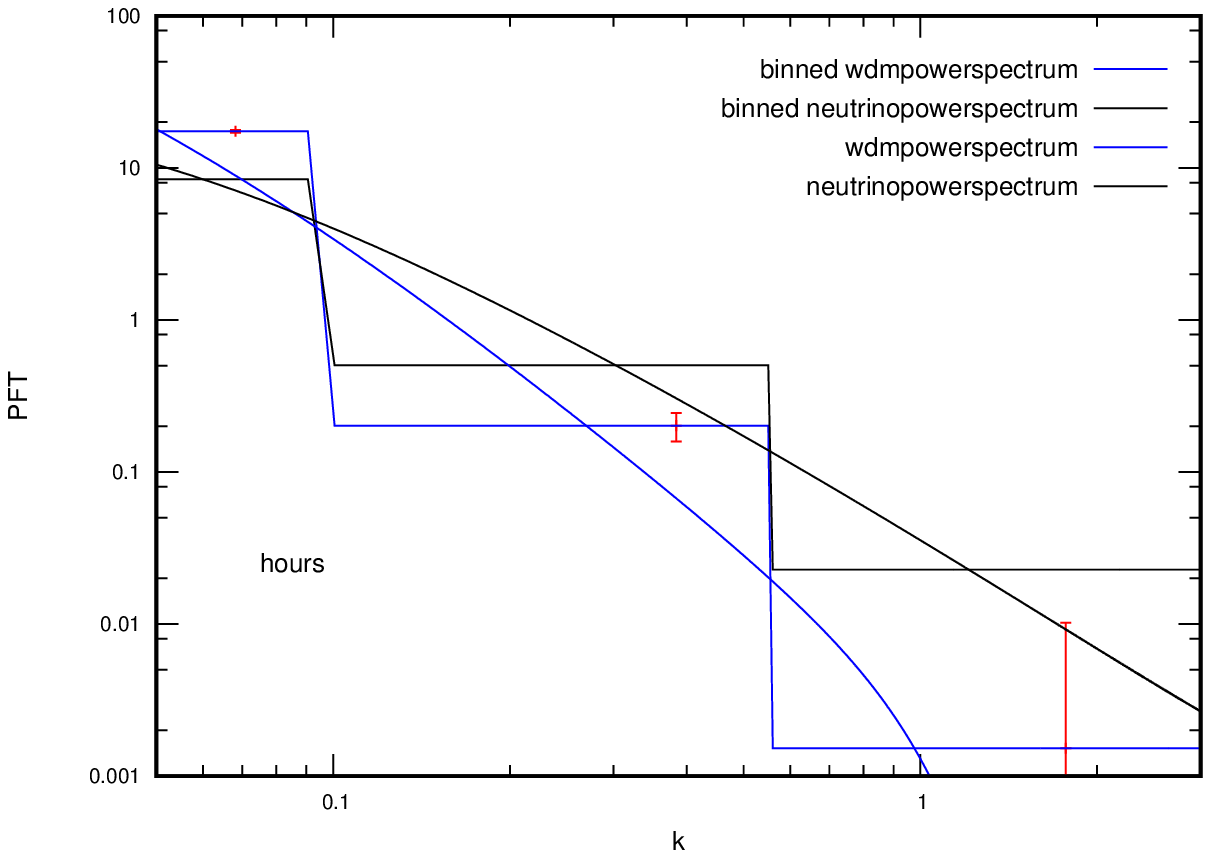}} {
    \includegraphics[scale =0.45]{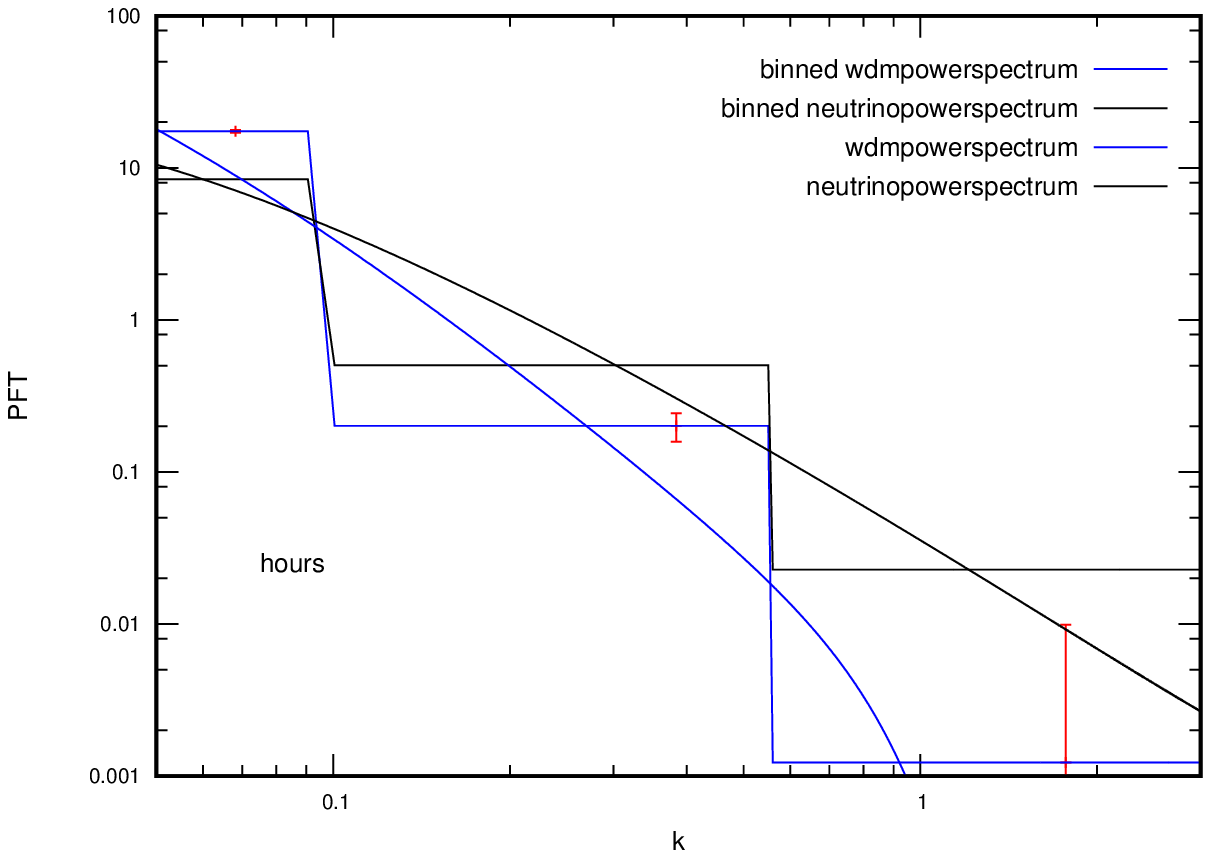}}}
\caption{Shows the $1-\sigma$ errors (in red) on the measurement of the WDM
  cross power spectrum $P_{\mathcal{F}T}^{\rm WDM} (k)$ in two different
  $k$-bins with an observing time of $20000$ hours distributed uniformly in
  $N_{\rm p}$ = 100 different fields-of-view. The left, central and right
  panels in the figure show the results for the cases of $m_{\rm WDM}$ = 0.20,
  0.15 and 0.10 keV respectively. The WDM (blue line) and the matter (black
  line) cross power spectrum are plotted for the binned WDM (blue
  horizontal steps) and CDM (black horizontal steps) cross power spectrum
  respectively.}
\label{fig:errorbin}
\end{center}
\end{figure}

As discussed earlier, we expect the WDM power spectrum to be
suppressed on small scales. We here focus on the possibility of measuring the suppression in the WDM power
spectrum in different $k$-bins using observations of the cross-correlation
signal of 200 hours each in 100 different fields-of-view. By measuring this
suppression with a high level of accuracy, one would be able to distinguish
between the WDM and CDM cross power spectrum. The suppression in the WDM
cross power spectrum is expected to be effective beyond a certain $k$, say
$k_{WDM}$. It is therefore meaningful to bin the $k$-modes with $k \geq
k_{WDM}$ in multiple $k$-bins and rest of the $k$-modes ($k < k_{WDM}$) in $k$-bin, and to consider measuring the WDM power spectrum in these
$k$-bins. The suppression is expected to be prominent at large $k$. One
may therefore hope to distinguish between the WDM and the CDM power
spectrum only if the error on the measurement of the WDM power spectrum in the bin corresponding to larger $k$
is small compared to the suppression in that $k$-bin. We see that for $m_{{\rm WDM}} = 0.25$ keV, $k_{WDM} \sim 0.1$ Mpc$^{-1}$. For the present purpose, we have divided the entires $k$-range into three different $k$-bins, $k$-modes with $k \leq 0.1$ Mpc$^{-1}$ are lumped into a single $k$-bin, and we have divided the rest of the $k$-range ($0.1 \leq k \leq 3.13$ Mpc$^{-1}$) into two equispaced logarithmic $k$-bins. The left panel of figure \ref{fig:errorbin} shows the predicted errors on the measurement of the WDM cross power spectrum for $m_{\rm WDM} = 0.25$ keV using 200 hours of observing time each in 100 different fields-of-view. We find that the errors on the measurement of the WDM power spectrum are $\sim 7$ and $\sim 2.5$ times smaller compared to the suppression of the WDM cross power spectrum in the two $k$-bins corresponding to the larger $k$-values. This refers to a measurement of the suppression at an confidence level of $6.9-\sigma$ and $2.5-\sigma$ respectively in two $k$-bins over the $k$-range $0.1 \leq k \leq 3.13$ Mpc$^{-1}$.

The suppression in the WDM power spectrum is relatively large for small
$m_{\rm WDM}$ as compared to the larger $m_{\rm WDM}$. This encourages us to
consider the possibility of improving the prospects of measuring the
suppression by lowering $m_{\rm WDM}$. We here note that the value of the
$k_{WDM}$ decreases as $m_{\rm WDM}$ is lowered. However, we have held the $k$-ranges 
corresponding to the three different $k$-bins fixed for the rest of our analysis. Ther predicted errors on the measurement of the WDM cross power spectrum is shown in the central and right panels of figure \ref{fig:errorbin} respectively for $m_{\rm WDM} = 0.20$ and $0.15$ keV. We find that errors decrease slightly, by factors of $1.01$ and $1.04$, and $1.04$ and $1.08$ in the two $k$-bins corresponding to the larger $k$-values respectively for $m_{\rm WDM} = 0.20$ and $0.15$ keV. This corresponds to measuring the suppression at confidence levels of $7-\sigma$ and $2.6-\sigma$, and $7.2-\sigma$ and $2.7-\sigma$ in these two $k$-bins over the $k$-range $0.1 \leq k \leq 3.13$ Mpc$^{-1}$ respectively for $m_{\rm WDM} = 0.20$ and $0.15$ keV.

\subsection{ Predictions from SKA I mid}
The baseline coverage of OWFA is small owing to the linear nature of the array. 
We now consider a radio-interferometric array for the 21-cm observation similar to the SKA1-mid. We have used the specifications of the radio telescope given in the 'Baseline Design Document'. We consider an interferometer with a total of $250$ antennae each of which has a diameter of $15$m. The range of operational frequencies is $350$MHz to $14$GHz. The baseline distribution of the array is obtained by assuming that $40\%$, $55\%$, $70\%$ and $100\%$ of the total number of antennae are within a  radius of $ 0.35$ km, $1$ km, $2.5$ km and $100$ km respectively. We also assume that below a radius of $30$m there is no baseline
For our analysis we have assumed a system temperature $T_{sys} = 70$K. We have also assumed the bandwidth of the telescope  to be $32$MHz and an average antenna efficiency of $0.7$
The formalism used in \cite{guha-sarkar2015} is used to compute the SNR and make Fisher matrix estimates for WDM mass.
\begin{figure}
\begin{center}
\includegraphics[scale=0.253]{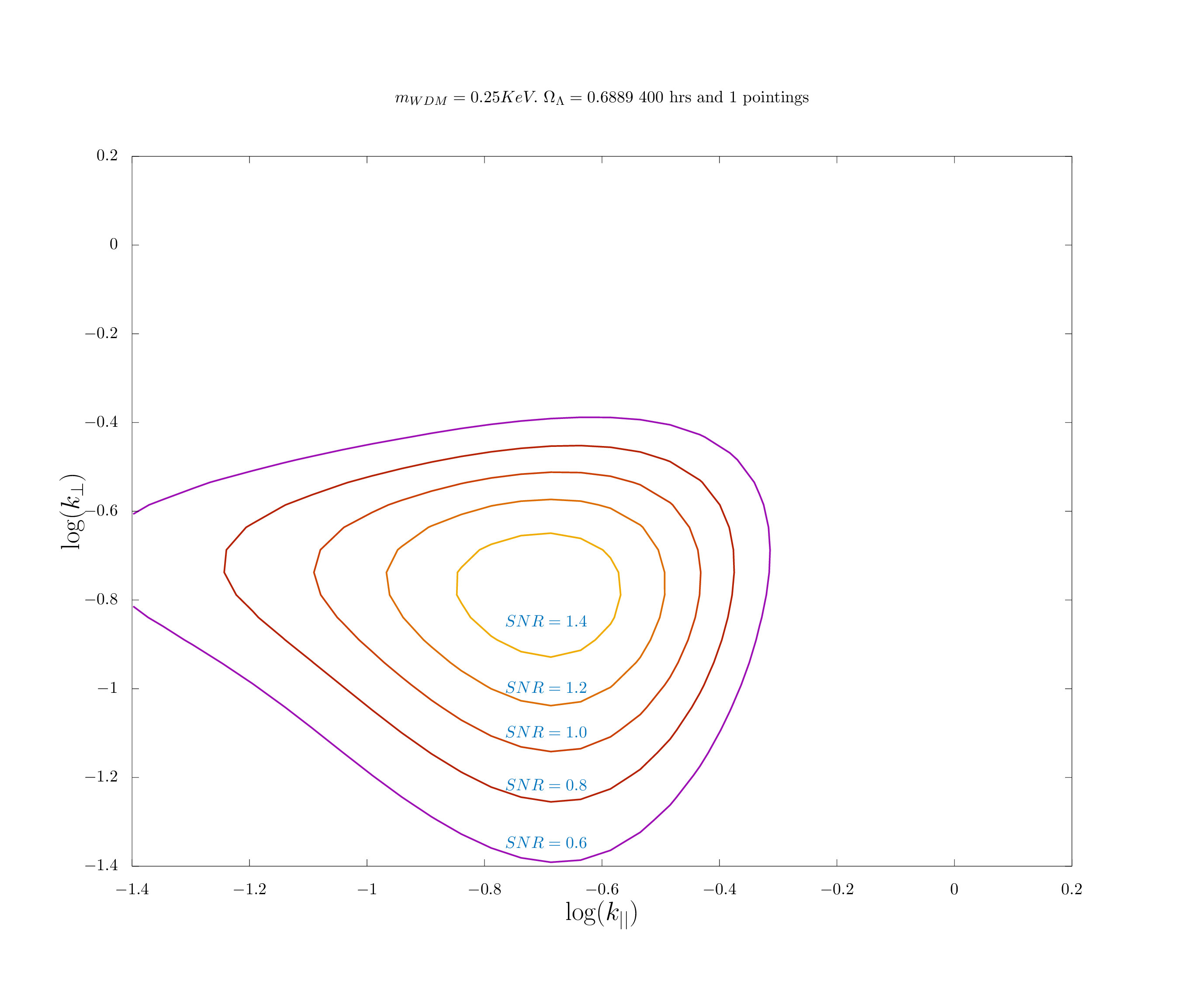}
\includegraphics[scale=0.253]{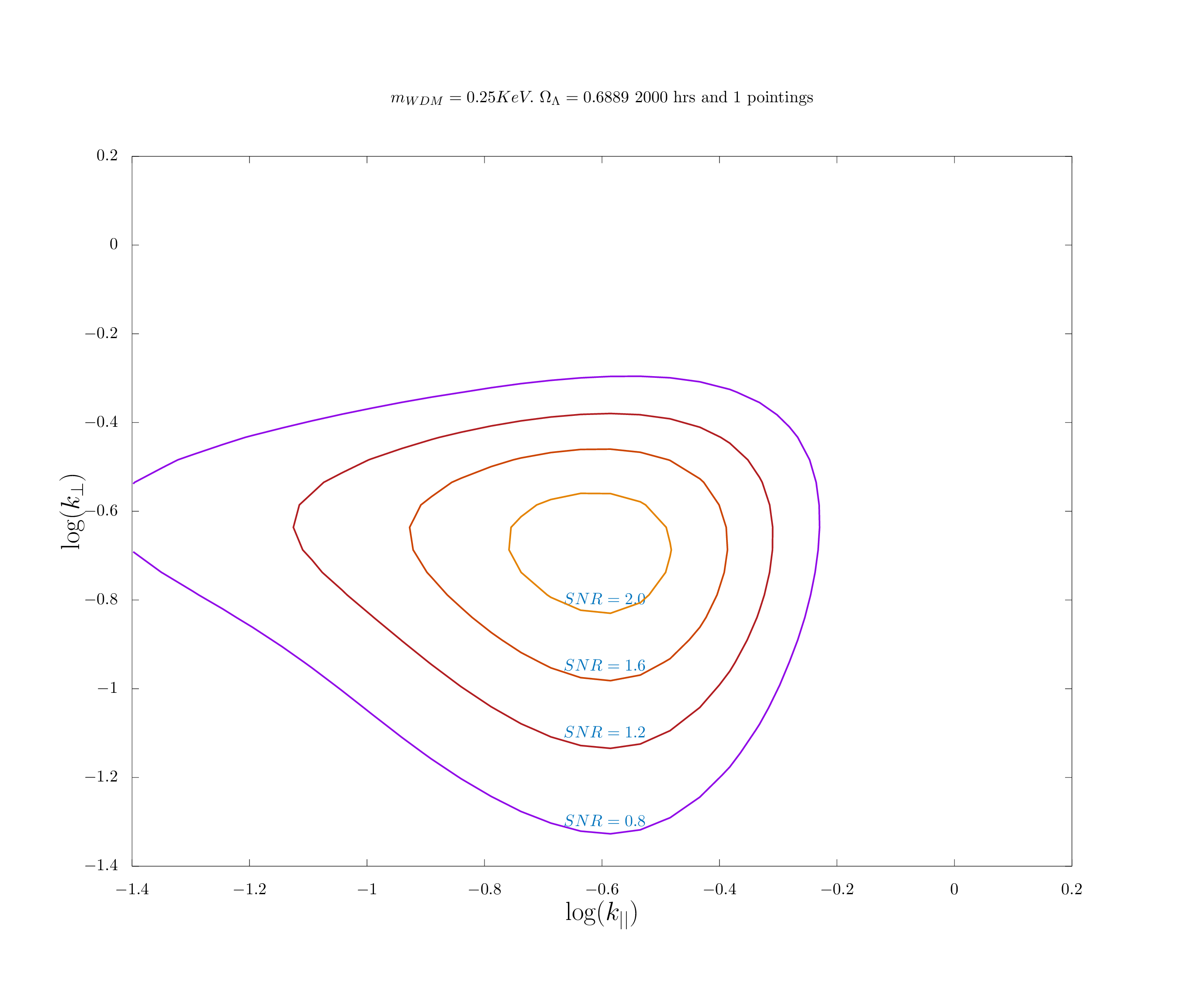}
\includegraphics[scale=0.253]{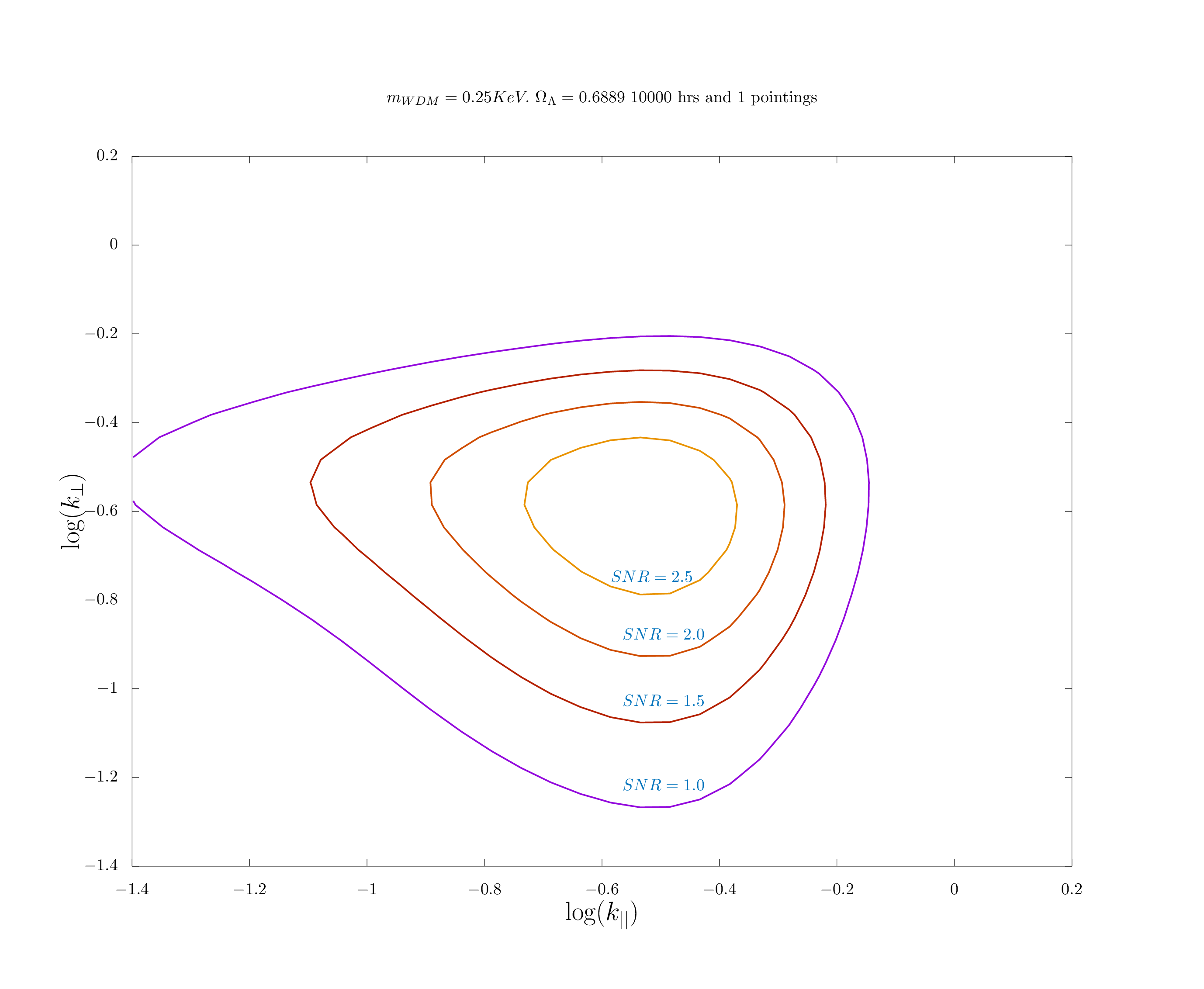}
\end{center}
\label{fig:ska1}
\caption{ Figure showing the SNR for measuring the suppression in  cross correlation power spectrum for a fiducial $m_{WDM} =0.25$Kev for a single pointing 21 cm observation for $400$hrs, $2000$hrs and $10000$hrs respectively.}
\end{figure}
For a single pointing observation of the cross-correlation signal with the instrument as  described above, the SNR improves very sluggishly beyond 400 hrs observations and saturates to a maximum of $3$ (peak in $k$-space) for a $20000$ hr observation which corresponds to the cosmic variance limit. Thus it is not worthwie to consider 20000hr observation in a single field of view. Prospects of higher SNR is possible if the observation time is divided over many fields of view. This also maximally utilizes Lyman-alpha spectra available for cross-correlation.

The figure \ref{fig:ska1} shows the SNR for the cross-correlation signal for a  fiducial WDM mass of  $m_{WDM} = 0.25 keV$ for observation time of $400$ hrs $2000$ hrs and $10000$ hrs in a single pointing observation.  The Lyman-alpha observational parameters are held fixed for this analysis. We find that if we consider a total observation time of $20000$ hrs divided over multiple pointings then a peak SNR of $\sqrt{50} \times 1.6  = 9.9$, $ \sqrt{10} \times 2.2 = 6.34$ and $\sqrt 2 \times 2.7 = 3.81$ is achievable in each of the cases respectively. The peak in SNR shifts in the $k$-space for different observation times  but it is in the typical range $0.1 \rm{Mpc}^{-1} < k < 0.39 \rm{Mpc}^{-1}$.

\begin{figure}

\centerline{
  {\includegraphics[scale = 0.75]{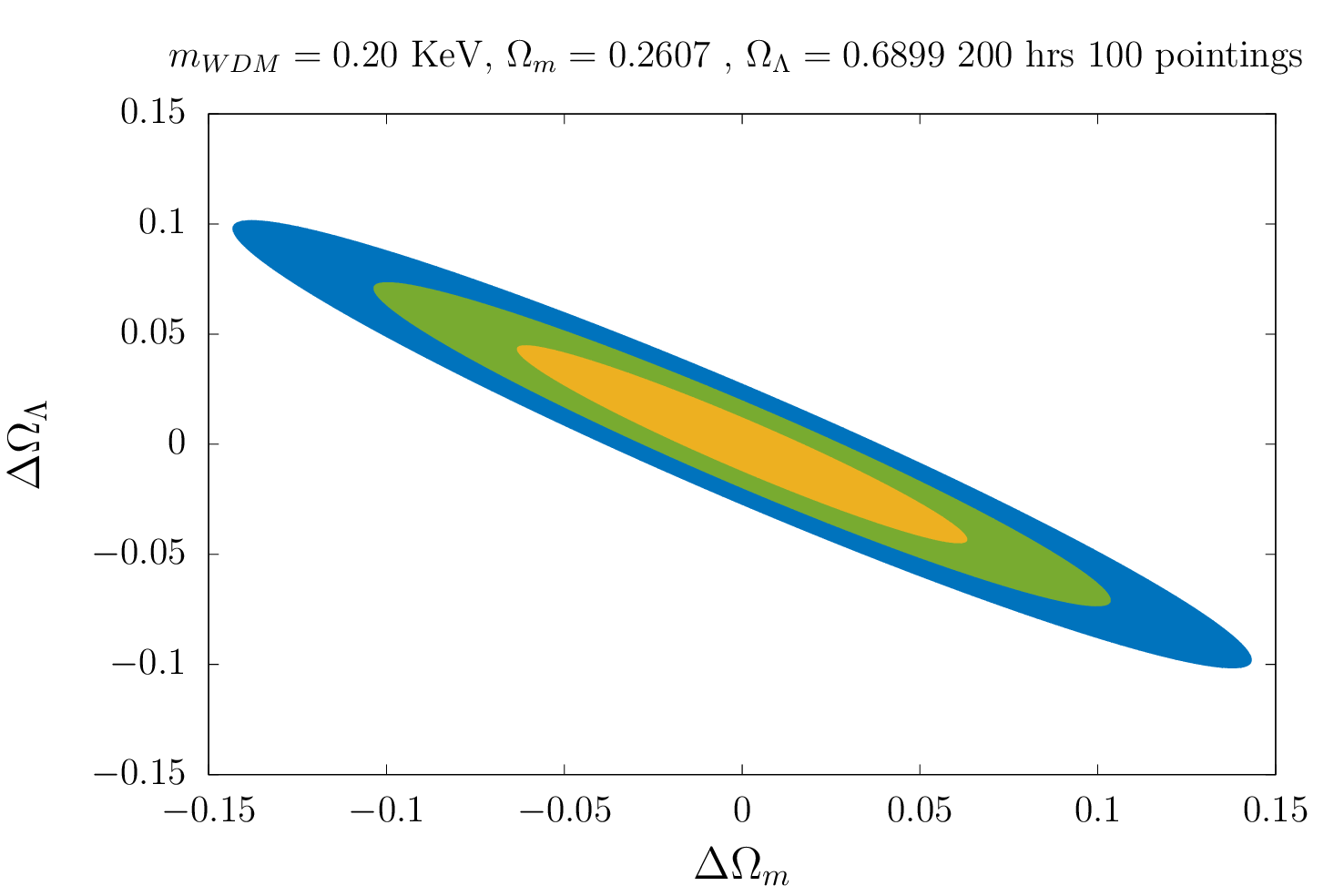}}}
\label{fig:skaellipse}
\caption {Figures showing the $68.3\%$, $ 95.4\%$ and $99.8\%$ error contours for $\Omega_{WDM}$ and $\Omega_{\Lambda}$ }
\end{figure}
The figure \ref{fig:skaellipse} shows the  $68.3\%$, $ 95.4\%$ and $99.8\%$ error contours for the joint estimation of $\Omega_{WDM}$ and $\Omega_{\Lambda}$. We have used the Fisher matrix formalism \cite {guha-sarkar2015} to obtain the error ellipse. We consider a 20000 hr observation in 100 pointings each of duration 200 hrs.
We find that for a fiducial WDM mass of $0.2$ keV, the 1-$\sigma$ relative error in $\Omega_{WDM}$ is  $\sim 0.16$. 
We also note that $\Omega_{\Lambda} $ is constrained at $\sim 7\%$.

\section{Conclusion}

In this article we have investigated the possibility of constraining WDM mass using the cross-correlation of the redshifted HI 21-cm signal and the Lyman-$\alpha$ forest from the post-reionization epoch. 
The effect of WDM on the suppression of the matter power spectrum has  the effect of enhancement of the 21-cm power spectrum through a non-linear bias  which is large on small scales owing to  the low abundance of small mass halos. We have considered a spectroscopic survey of the quasars with a mean quasar number density of $\bar{{\rm n}}_{\rm Q} = 48$ deg$^{-2}$ over a survey area of 14455 deg$^{2}$, and  a mean spectroscopic SNR = 5. 
We have chosen two futuristic radio-interferometers for our cross-correlation analysis namely OWFA and SKA1-mid. These two telescopes differ in their array layout and thereby in their baseline distribution. The former is a linear one-dimensional array and the latter is laid out in two dimensions. 

Our analysis with OWFA shows that it is possible to measure the WDM power spectrum in several $k$-bins with ${\rm SNR} > 5$ using an observation of 200 hours each in 100 different fields-of-view for $m_{\rm WDM} = 0.25$ keV. The relative $1-\sigma$ error in measurement of the parameter $\Omega_{{\rm WDM}}$ is $\sim 0.8$ for a fiducial $m_{\rm WDM} = 0.25$ keV. It is also possible to have a measurement of the suppression of power from the Cold Dark Matter (CDM) power spectrum at a confidence level of $\sim 7.2-\sigma$ and $\sim 2.7-\sigma$ in two different $k$-bins over the $k$-range $0.1 \leq k \leq 3.13$ Mpc$^{-1}$ for $m_{\rm WDM} = 0.15$ keV. Considering a cross-correlation  with SKA1-mid, we find that for a fiducial $m_{\rm WDM}= 0.25$ Kev, the  suppression in the cross  power spectrum can be measured at $\sim 10 - \sigma$ around $k \sim 0.2 \rm{Mpc}^{-1}$ for a total observing time of $20000$ hrs distributed uniformly over $50$ independent pointings. In summary, our study indicates that the cross-correlation of Lyman-$\alpha$ and post-reionization 21 cm signal maybe effective in putting cosmological bounds on WDM theories with far less severity of observational issues like foreground subtraction and systematics.

\bibliographystyle{JHEP} 
\bibliography{ref}

\providecommand{\href}[2]{#2}\begingroup\raggedright\begin{thebibliography}{10}

\bibitem{aghanim2018planck}
N.~Aghanim, Y.~Akrami, M.~Ashdown, J.~Aumont, C.~Baccigalupi, M.~Ballardini
  et~al., \emph{Planck 2018 results. vi. cosmological parameters}, {\emph{arXiv
  preprint arXiv:1807.06209} (2018) }.

\bibitem{SDSS14}
B.~Abolfathi, D.~Aguado, G.~Aguilar, C.~A. Prieto, A.~Almeida, T.~T. Ananna
  et~al., \emph{The fourteenth data release of the sloan digital sky survey:
  first spectroscopic data from the extended baryon oscillation sky survey and
  from the second phase of the apache point observatory galactic evolution
  experiment}, {\emph{arXiv preprint arXiv:1707.09322} (2017) }.

\bibitem{moore1999dark}
B.~Moore, S.~Ghigna, F.~Governato, G.~Lake, T.~Quinn, J.~Stadel et~al.,
  \emph{Dark matter substructure within galactic halos}, {\emph{The
  Astrophysical Journal Letters} {\bfseries 524} (1999) L19}.

\bibitem{klypin1999missing}
A.~Klypin, A.~V. Kravtsov, O.~Valenzuela and F.~Prada, \emph{Where are the
  missing galactic satellites?}, {\emph{The Astrophysical Journal} {\bfseries
  522} (1999) 82}.

\bibitem{peebles2010nearby}
P.~Peebles and A.~Nusser, \emph{Nearby galaxies as pointers to a better theory
  of cosmic evolution}, {\emph{Nature} {\bfseries 465} (2010) 565}.

\bibitem{diemand2007formation}
J.~Diemand, M.~Kuhlen and P.~Madau, \emph{Formation and evolution of galaxy
  dark matter halos and their substructure}, {\emph{The Astrophysical Journal}
  {\bfseries 667} (2007) 859}.

\bibitem{navarro1997universal}
J.~F. Navarro, C.~S. Frenk and S.~D. White, \emph{A universal density profile
  from hierarchical clustering}, {\emph{The Astrophysical Journal} {\bfseries
  490} (1997) 493}.

\bibitem{stadel2009quantifying}
J.~Stadel, D.~Potter, B.~Moore, J.~Diemand, P.~Madau, M.~Zemp et~al.,
  \emph{Quantifying the heart of darkness with ghalo--a multibillion particle
  simulation of a galactic halo}, {\emph{Monthly Notices of the Royal
  Astronomical Society: Letters} {\bfseries 398} (2009) L21--L25}.

\bibitem{boyarsky2009lyman}
A.~Boyarsky, J.~Lesgourgues, O.~Ruchayskiy and M.~Viel, \emph{Lyman-$\alpha$
  constraints on warm and on warm-plus-cold dark matter models}, {\emph{Journal
  of Cosmology and Astroparticle Physics} {\bfseries 2009} (2009) 012}.

\bibitem{boyarsky2008constraints}
A.~Boyarsky, D.~Iakubovskyi, O.~Ruchayskiy and V.~Savchenko, \emph{Constraints
  on decaying dark matter from xmm--newton observations of m31}, {\emph{Monthly
  Notices of the Royal Astronomical Society} {\bfseries 387} (2008)
  1361--1373}.

\bibitem{boyarsky2009realistic}
A.~Boyarsky, J.~Lesgourgues, O.~Ruchayskiy and M.~Viel, \emph{Realistic sterile
  neutrino dark matter with kev mass does not contradict cosmological bounds},
  {\emph{Physical review letters} {\bfseries 102} (2009) 201304}.

\bibitem{seljak2006can}
U.~Seljak, A.~Makarov, P.~McDonald and H.~Trac, \emph{Can sterile neutrinos be
  the dark matter?}, {\emph{Physical Review Letters} {\bfseries 97} (2006)
  191303}.

\bibitem{ellis1984supersymmetric}
J.~Ellis, J.~S. Hagelin, D.~V. Nanopoulos, K.~Olive and M.~Srednicki,
  \emph{Supersymmetric relics from the big bang}, {\emph{Nuclear Physics B}
  {\bfseries 238} (1984) 453--476}.

\bibitem{dodelson1994sterile}
S.~Dodelson and L.~M. Widrow, \emph{Sterile neutrinos as dark matter},
  {\emph{Physical Review Letters} {\bfseries 72} (1994) 17}.

\bibitem{bode2001halo}
P.~Bode, J.~P. Ostriker and N.~Turok, \emph{Halo formation in warm dark matter
  models}, {\emph{The Astrophysical Journal} {\bfseries 556} (2001) 93}.

\bibitem{boyanovsky2011warm}
D.~Boyanovsky, \emph{Warm dark matter at small scales: peculiar velocities and
  phase space density}, {\emph{Physical Review D} {\bfseries 83} (2011)
  103504}.

\bibitem{BNS2001}
S.~Bharadwaj, B.~B. Nath and S.~K. Sethi, \emph{Using hi to probe large scale
  structures at z ~ 3}, {\emph{Journal of Astrophysics and Astronomy}
  {\bfseries 22} (2001) 21--34}.

\bibitem{bharadwaj-sethi2001}
S.~Bharadwaj and S.~K. Sethi, \emph{Hi fluctuations at large redshifts:
  I-visibility correlation}, {\emph{Journal of Astrophysics and Astronomy}
  {\bfseries 22} (2001) 293--307}.

\bibitem{wyithe-loeb2008}
A.~Loeb and J.~S.~B. Wyithe, \emph{Possibility of precise measurement of the
  cosmological power spectrum with a dedicated survey of 21 cm emission after
  reionization}, {\emph{Physical Review Letters} {\bfseries 100} (2008)
  161301}.

\bibitem{bharadwaj2009}
S.~Bharadwaj, S.~K. Sethi and T.~D. Saini, \emph{Estimation of cosmological
  parameters from neutral hydrogen observations of the post-reionization
  epoch}, {\emph{Physical Review D} {\bfseries 79} (2009) 083538}.

\bibitem{visbal2009}
E.~Visbal, A.~Loeb and S.~Wyithe, \emph{Cosmological constraints from 21cm
  surveys after reionization}, {\emph{Journal of Cosmology and Astroparticle
  Physics} {\bfseries 2009} (2009) 030}.

\bibitem{ananthakrishnan1995GMRT}
S.~Ananthakrishnan, \emph{The giant meterwave radio telescope/gmrt},
  {\emph{Journal of Astrophysics and Astronomy Supplement} {\bfseries 16}
  (1995) 427}.

\bibitem{prasad2011}
P.~Prasad and C.~Subrahmanya, \emph{A high speed networked signal processing
  platform for multi-element radio telescopes}, {\emph{Experimental Astronomy}
  {\bfseries 31} (2011) 1--22}.

\bibitem{subrahmanya2017OWFA}
C.~Subrahmanya, P.~Manoharan and J.~N. Chengalur, \emph{The ooty wide field
  array}, {\emph{Journal of Astrophysics and Astronomy} {\bfseries 38} (2017)
  10}.

\bibitem{bandura2014CHIME}
K.~Bandura, G.~E. Addison, M.~Amiri, J.~R. Bond, D.~Campbell-Wilson, L.~Connor
  et~al., \emph{Canadian hydrogen intensity mapping experiment (chime)
  pathfinder},  in \emph{Ground-based and Airborne Telescopes V}, vol.~9145,
  p.~914522, International Society for Optics and Photonics, 2014.

\bibitem{ghosh2011}
A.~Ghosh, S.~Bharadwaj, S.~S. Ali and J.~N. Chengalur, \emph{Improved
  foreground removal in gmrt 610 mhz observations towards redshifted 21-cm
  tomography}, {\emph{Monthly Notices of the Royal Astronomical Society}
  {\bfseries 418} (2011) 2584--2589}.

\bibitem{ghosh2012foregrounds}
A.~Ghosh, J.~Prasad, S.~Bharadwaj, S.~S. Ali and J.~N. Chengalur,
  \emph{Characterizing foreground for redshifted 21 cm radiation: 150 mhz giant
  metrewave radio telescope observations}, {\emph{Monthly Notices of the Royal
  Astronomical Society} {\bfseries 426} (2012) 3295--3314}.

\bibitem{prasad2013foregrounds}
J.~Prasad, A.~Ghosh, S.~Bharadwaj, J.~N. Chengalur and S.~Ali,
  \emph{Characterizing foreground for redshifted 21-cm radiation: 150 mhz gmrt
  observations}, .

\bibitem{ali-bharadwaj2014}
S.~S. Ali and S.~Bharadwaj, \emph{Prospects for detecting the 326.5 mhz
  redshifted 21-cm hi signal with the ooty radio telescope (ort)},
  {\emph{Journal of Astrophysics and Astronomy} {\bfseries 35} (2014)
  157--182}.

\bibitem{croft-weinberg1998lyman}
R.~A. Croft, D.~H. Weinberg, N.~Katz and L.~Hernquist, \emph{Recovery of the
  power spectrum of mass fluctuations from observations of the ly$\alpha$
  forest}, {\emph{The Astrophysical Journal} {\bfseries 495} (1998) 44}.

\bibitem{croft-weinberg1999lyman}
R.~A. Croft, D.~H. Weinberg, M.~Pettini, L.~Hernquist and N.~Katz, \emph{The
  power spectrum of mass fluctuations measured from the ly$\alpha$ forest at
  redshift z= 2.5}, {\emph{The Astrophysical Journal} {\bfseries 520} (1999)
  1}.

\bibitem{croft-weinberg2002lyman}
R.~A. Croft, D.~H. Weinberg, M.~Bolte, S.~Burles, L.~Hernquist, N.~Katz et~al.,
  \emph{Toward a precise measurement of matter clustering: Ly$\alpha$ forest
  data at redshifts 2-4}, {\emph{The Astrophysical Journal} {\bfseries 581}
  (2002) 20}.

\bibitem{mandelbaum2003lyman}
R.~Mandelbaum, P.~McDonald, U.~Seljak and R.~Cen, \emph{Precision cosmology
  from the lyman $\alpha$ forest: power spectrum and bispectrum},
  {\emph{Monthly Notices of the Royal Astronomical Society} {\bfseries 344}
  (2003) 776--788}.

\bibitem{vielbispectrum2004lyman}
M.~Viel, S.~Matarrese, A.~Heavens, M.~Haehnelt, T.-S. Kim, V.~Springel et~al.,
  \emph{The bispectrum of the lyman $\alpha$ forest at z< 2-2.4 from a large
  sample of uves qso absorption spectra (luqas)}, {\emph{Monthly Notices of the
  Royal Astronomical Society} {\bfseries 347} (2004) L26--L30}.

\bibitem{mcdonald1999lyman}
P.~McDonald and J.~Miralda-Escud{\'e}, \emph{Measuring the cosmological
  geometry from the ly$\alpha$ forest along parallel lines of sight},
  {\emph{The Astrophysical Journal} {\bfseries 518} (1999) 24}.

\bibitem{lesgourgues2007lyman}
J.~Lesgourgues, M.~Viel, M.~Haehnelt and R.~Massey, \emph{A combined analysis
  of lyman-$\alpha$ forest, 3 d weak lensing and wmap year three data j.
  cosmol}, {\emph{Astropart. Phys. JCAP11 (2007)} {\bfseries 8} (2007) }.

\bibitem{gallerani2006reionization}
S.~Gallerani, T.~R. Choudhury and A.~Ferrara, \emph{Constraining the
  reionization history with qso absorption spectra}, {\emph{Monthly Notices of
  the Royal Astronomical Society} {\bfseries 370} (2006) 1401--1421}.

\bibitem{mcdonald2007lyman}
P.~McDonald and D.~J. Eisenstein, \emph{Dark energy and curvature from a future
  baryonic acoustic oscillation survey using the lyman-$\alpha$ forest},
  {\emph{Physical Review D} {\bfseries 76} (2007) 063009}.

\bibitem{hui-gnedin1997}
L.~Hui and N.~Y. Gnedin, \emph{Equation of state of the photoionized
  intergalactic medium}, {\emph{Monthly Notices of the Royal Astronomical
  Society} {\bfseries 292} (1997) 27--42}.

\bibitem{weinberg1997lymana}
D.~H. Weinberg, L.~Hernquist, N.~Katz, R.~Croft and J.~Miralda-Escud{\'e},
  \emph{Hubble flow broadening of the lyman-alpha forest and its implications},
  {\emph{arXiv preprint astro-ph/9709303} (1997) }.

\bibitem{weinberg1997lymanb}
D.~H. Weinberg, L.~Hernquist and N.~Katz, \emph{Photoionization, numerical
  resolution, and galaxy formation}, {\emph{The Astrophysical Journal}
  {\bfseries 477} (1997) 8}.

\bibitem{kim2007metal}
T.-S. Kim, J.~Bolton, M.~Viel, M.~Haehnelt and R.~Carswell, \emph{An improved
  measurement of the flux distribution of the ly$\alpha$ forest in qso
  absorption spectra: the effect of continuum fitting, metal contamination and
  noise properties}, {\emph{Monthly Notices of the Royal Astronomical Society}
  {\bfseries 382} (2007) 1657--1674}.

\bibitem{bagla2010}
J.~Bagla, N.~Khandai and K.~K. Datta, \emph{Hi as a probe of the large-scale
  structure in the post-reionization universe}, {\emph{Monthly Notices of the
  Royal Astronomical Society} {\bfseries 407} (2010) 567--580}.

\bibitem{guha-sarkar2012}
T.~Guha~Sarkar, S.~Mitra, S.~Majumdar and T.~R. Choudhury, \emph{Constraining
  large-scale hi bias using redshifted 21-cm signal from the post-reionization
  epoch}, {\emph{Monthly Notices of the Royal Astronomical Society} {\bfseries
  421} (2012) 3570--3578}.

\bibitem{sarkar2016}
D.~Sarkar, S.~Bharadwaj and S.~Anathpindika, \emph{Modelling the
  post-reionization neutral hydrogen (hi) bias}, {\emph{Monthly Notices of the
  Royal Astronomical Society} {\bfseries 460} (2016) 4310--4319}.

\bibitem{carucci2017}
I.~P. Carucci, F.~Villaescusa-Navarro and M.~Viel, \emph{The cross-correlation
  between 21 cm intensity mapping maps and the ly$\alpha$ forest in the
  post-reionization era}, {\emph{Journal of Cosmology and Astroparticle
  Physics} {\bfseries 2017} (2017) 001}.

\bibitem{Guha-sarkar2011cross}
T.~Guha~Sarkar, S.~Bharadwaj, T.~R. Choudhury and K.~K. Datta,
  \emph{Cross-correlation of the hi 21-cm signal and lyα forest: a probe of
  cosmology}, {\emph{Monthly Notices of the Royal Astronomical Society}
  {\bfseries 410} (2011) 1130--1134}.

\bibitem{guha-sarkar2015}
T.~G. Sarkar and K.~K. Datta, \emph{On using large scale correlation of the
  ly-$\alpha$ forest and redshifted 21-cm signal to probe hi distribution
  during the post reionization era}, {\emph{Journal of Cosmology and
  Astroparticle Physics} {\bfseries 2015} (2015) 001}.

\bibitem{pal2016}
A.~K. Pal and T.~Guha~Sarkar, \emph{Constraining neutrino mass using the
  large-scale h i distribution in the post-reionization epoch}, {\emph{Monthly
  Notices of the Royal Astronomical Society} {\bfseries 459} (2016)
  3505--3511}.

\bibitem{guha-sarkaretal2016}
T.~G. Sarkar, K.~Datta, A.~Pal, T.~R. Choudhury and S.~Bharadwaj,
  \emph{Redshifted hi 21-cm signal from the post-reionization epoch:
  Cross-correlations with other cosmological probes}, {\emph{Journal of
  Astrophysics and Astronomy} {\bfseries 37} (2016) 26}.

\bibitem{guha-sarkarsen2016}
T.~G. Sarkar and A.~A. Sen, \emph{Cosmology and astrophysics using the
  post-reionization hi}, {\emph{Journal of Astrophysics and Astronomy}
  {\bfseries 37} (2016) 33}.

\bibitem{sarkar2018cross}
A.~K. Sarkar, S.~Bharadwaj and T.~G. Sarkar, \emph{Predictions for measuring
  the cross power spectrum of the hi 21-cm signal and the lyman-$\alpha$ forest
  using owfa}, {\emph{Journal of Cosmology and Astroparticle Physics}
  {\bfseries 2018} (2018) 051}.

\bibitem{villaescusa2015lymanbreak}
F.~Villaescusa-Navarro, M.~Viel, D.~Alonso, K.~K. Datta, P.~Bull and M.~G.
  Santos, \emph{Cross-correlating 21cm intensity maps with lyman break galaxies
  in the post-reionization era}, {\emph{Journal of Cosmology and Astroparticle
  Physics} {\bfseries 2015} (2015) 034}.

\bibitem{chang2010HIintensity}
T.-C. Chang, U.-L. Pen, K.~Bandura and J.~B. Peterson, \emph{An intensity map
  of hydrogen 21-cm emission at redshift z≈ 0.8}, {\emph{Nature} {\bfseries
  466} (2010) 463}.

\bibitem{smith2011WDM}
R.~E. Smith and K.~Markovic, \emph{Testing the warm dark matter paradigm with
  large-scale structures}, {\emph{Physical Review D} {\bfseries 84} (2011)
  063507}.

\bibitem{viel2005constraining}
M.~Viel, J.~Lesgourgues, M.~G. Haehnelt, S.~Matarrese and A.~Riotto,
  \emph{Constraining warm dark matter candidates including sterile neutrinos
  and light gravitinos with wmap and the lyman-$\alpha$ forest},
  {\emph{Physical Review D} {\bfseries 71} (2005) 063534}.

\bibitem{eisenstein-hu1998}
D.~J. Eisenstein and W.~Hu, \emph{Baryonic features in the matter transfer
  function}, {\emph{The Astrophysical Journal} {\bfseries 496} (1998) 605}.

\bibitem{loeb2001reionization}
A.~Loeb and R.~Barkana, \emph{The reionization of the universe by the first
  stars and quasars}, {\emph{Annual review of astronomy and astrophysics}
  {\bfseries 39} (2001) 19--66}.

\bibitem{barkana2001reionization}
R.~Barkana and A.~Loeb, \emph{In the beginning: the first sources of light and
  the reionization of the universe}, {\emph{Physics reports} {\bfseries 349}
  (2001) 125--238}.

\bibitem{mesinger2015reionization}
A.~Mesinger, \emph{Understanding the Epoch of Cosmic Reionization: Challenges
  and Progress}, vol.~423.
\newblock Springer, 2015.

\bibitem{bharadwaj-ali2004CMB}
S.~Bharadwaj and S.~S. Ali, \emph{The cosmic microwave background radiation
  fluctuations from h i perturbations prior to reionization}, {\emph{Monthly
  Notices of the Royal Astronomical Society} {\bfseries 352} (2004) 142--146}.

\bibitem{pontzen2008lyman}
A.~Pontzen, F.~Governato, M.~Pettini, C.~Booth, G.~Stinson, J.~Wadsley et~al.,
  \emph{Damped lyman $\alpha$ systems in galaxy formation simulations},
  {\emph{Monthly Notices of the Royal Astronomical Society} {\bfseries 390}
  (2008) 1349--1371}.

\bibitem{palanque2013}
N.~Palanque-Delabrouille, C.~Y{\`e}che, A.~Borde, J.-M. Le~Goff, G.~Rossi,
  M.~Viel et~al., \emph{The one-dimensional ly$\alpha$ forest power spectrum
  from boss}, {\emph{Astronomy \& Astrophysics} {\bfseries 559} (2013) A85}.

\bibitem{bharadwaj-ali2005}
S.~Bharadwaj and S.~Saiyad~Ali, \emph{On using visibility correlations to probe
  the hi distribution from the dark ages to the present epoch--i. formalism and
  the expected signal}, {\emph{Monthly Notices of the Royal Astronomical
  Society} {\bfseries 356} (2005) 1519--1528}.

\bibitem{zafar2013}
T.~Zafar, A.~Popping and C.~P{\'e}roux, \emph{The eso uves advanced data
  products quasar sample-i. dataset and new $n_{HI}$ measurements of damped
  absorbers}, {\emph{Astronomy \& Astrophysics} {\bfseries 556} (2013) A140}.

\bibitem{prochaska2009}
J.~X. Prochaska and A.~M. Wolfe, \emph{On the (non) evolution of hi gas in
  galaxies over cosmic time}, {\emph{The Astrophysical Journal} {\bfseries 696}
  (2009) 1543}.

\bibitem{noterdaeme2012}
P.~Noterdaeme, P.~Petitjean, W.~Carithers, I.~P{\^a}ris, A.~Font-Ribera,
  S.~Bailey et~al., \emph{Column density distribution and cosmological mass
  density of neutral gas: Sloan digital sky survey-iii data release 9},
  {\emph{Astronomy \& Astrophysics} {\bfseries 547} (2012) L1}.

\bibitem{villaescusa2014modelling}
F.~Villaescusa-Navarro, M.~Viel, K.~K. Datta and T.~R. Choudhury,
  \emph{Modeling the neutral hydrogen distribution in the post-reionization
  universe: intensity mapping}, {\emph{Journal of Cosmology and Astroparticle
  Physics} {\bfseries 2014} (2014) 050}.

\bibitem{marthi2013}
V.~R. Marthi and J.~Chengalur, \emph{Non-linear redundancy calibration},
  {\emph{Monthly Notices of the Royal Astronomical Society} {\bfseries 437}
  (2013) 524--531}.

\bibitem{swarup1971}
G.~Swarup, N.~Sarma, M.~Joshi, V.~Kapahi, D.~Bagri, S.~Damle et~al.,
  \emph{Large steerable radio telescope at ootacamund, india}, {\emph{Nature
  Physical Science} {\bfseries 230} (1971) 185}.

\bibitem{sarma1975}
N.~Sarma, M.~Joshi, D.~Bagri and S.~Ananthakrishnan, \emph{Receiver system of
  the ooty radio telescope}, {\emph{IETE Journal of Research} {\bfseries 21}
  (1975) 110--116}.

\bibitem{gehlot-bagla2017OWFA}
B.~K. Gehlot and J.~S. Bagla, \emph{{Prospects of detecting HI using redshifted
  21-cm radiation at $z \sim 3$}}, {\emph{Journal of Astrophysics and
  Astronomy} {\bfseries 38} (2017) 13}.

\bibitem{marthi2017OWFA}
V.~R. Marthi, \emph{Prowess--a software model for the ooty wide field array},
  {\emph{Journal of Astrophysics and Astronomy} {\bfseries 38} (2017) 12}.

\bibitem{chatterjee2017OWFA}
S.~Chatterjee, S.~Bharadwaj and V.~R. Marthi, \emph{{Simulating the z=3.35 HI
  21-cm visibility signal for the Ooty Wide Field Array (OWFA)}},
  {\emph{Journal of Astrophysics and Astronomy} {\bfseries 38} (2017) 15}.

\bibitem{chatterjee2018OWFA}
S.~Chatterjee and S.~Bharadwaj, \emph{A spherical harmonic analysis of the ooty
  wide field array (owfa) visibility signal}, {\emph{Monthly Notices of the
  Royal Astronomical Society} (2018) }.

\bibitem{marthi-chatterjee2017OWFA}
V.~R. Marthi, S.~Chatterjee, J.~N. Chengalur and S.~Bharadwaj, \emph{{Simulated
  predictions for HI at z= 3.35 with the Ooty Wide Field Array--I. Instrument
  and the foregrounds}}, {\emph{Monthly Notices of the Royal Astronomical
  Society} {\bfseries 471} (2017) 3112--3126}.

\bibitem{marthi-chengalur2013OWFA}
V.~R. Marthi and J.~Chengalur, \emph{Non-linear redundancy calibration},
  {\emph{Monthly Notices of the Royal Astronomical Society} {\bfseries 437}
  (2013) 524--531}.

\bibitem{}
R.~Mandelbaum, P.~McDonald, U.~Seljak and R.~Cen, \emph{Precision cosmology
  from the lyman $\alpha$ forest: power spectrum and bispectrum},
  {\emph{Monthly Notices of the Royal Astronomical Society} {\bfseries 344}
  (2003) 776--788}.

\end{thebibliography}\endgroup

\end{document}